%% file: document.tex
\documentclass[11pt]{article}
\usepackage[margin=1in]{geometry}
\usepackage{microtype}
\usepackage{amsmath}
\usepackage{graphicx}
\usepackage{amssymb}
\usepackage{amsthm}
\usepackage{verbatim}
\usepackage{epsfig}
\usepackage{color}
\usepackage{url}
\usepackage{listings}
\usepackage{float}
\usepackage[caption=false]{subfig}
\usepackage{cite} 
\usepackage{authblk} 
\usepackage{mathtools}
\usepackage{tcolorbox}
\usepackage{nicefrac}
\usepackage{booktabs}
\usepackage[pdftex, plainpages = false, pdfpagelabels, 
				bookmarks=false,
				bookmarksopen = true,
				bookmarksnumbered = true,
				breaklinks = true,
				linktocpage,
				pagebackref,
				colorlinks = true,  
				linkcolor = blue,
				urlcolor  = blue,
				citecolor = red,
				anchorcolor = green,
				hyperindex = true,
				hyperfigures
]{hyperref}

\newtheorem{lemma}{Lemma}
\newtheorem{theorem}{Theorem}

\usepackage{amsmath,amssymb,latexsym,epsfig,amsthm,yhmath}
\usepackage[margin=1in]{geometry}
\usepackage{amssymb,latexsym,amsthm}
\usepackage{adjustbox}
\usepackage{algorithm}
\usepackage{appendix}
\usepackage{etoolbox}
\usepackage{xspace} 
\usepackage{color}%
\usepackage[font={small, it}]{caption}
\usepackage{wrapfig}
\usepackage{thmtools, thm-restate} 
\usepackage[para]{footmisc}

\newtheorem{remark}{Remark}
\newtheorem{corollary}[theorem]{Corollary}

\usepackage{algorithm}
\usepackage{xspace}
\long\def\cut#1{{}}
\def\cc{\mathcal{C}}
\def\cp{\mathcal{P}}
\def\ca{\mathcal{A}}

\def\cs{\mathcal{S}}
\def\cg{\mathcal{G}}

\def\rs{\mathcal{R}} 
\def\closer{\textit{closer}\xspace}
\def\farther{\textit{farther}\xspace}
\def\tz{\text{Type-0}\xspace}
\def\to{\text{Type-1}\xspace}
\def\pbf{q_0^*}
\def\ptf{q_1^*}
\def\rbf{Q_0^*}
\def\rtf{Q_1^*}
\def\pvec{\mathbf{q}}
\def\rvec{\mathbf{Q}}
\def\pvecf{\mathbf{q^*}}
\def\rvecf{\mathbf{Q^*}}

\newcommand{\np}{\textsf{NP}}
\def\eps{\epsilon}

\def\mathify#1{\ifmmode{#1}\else\mbox{$#1$}\fi} 

\title{The Maximum Exposure Problem}
\author[1]{Neeraj Kumar\thanks{neeraj@cs.ucsb.edu}}
\author[2]{Stavros Sintos\thanks{ssintos@cs.duke.edu}}
\author[1]{Subhash Suri\thanks{suri@cs.ucsb.edu}}
\affil[1]{Department of Computer Science, University of California, Santa Barbara, USA}
\affil[2]{Duke University, Durham, NC, USA}
\date{}

\begin{document}
\maketitle
\begin{abstract}
	Given a set of points $P$ and axis-aligned rectangles $\rs$ in the plane, a point $p \in P$
	is called \emph{exposed} if it lies outside all rectangles in $\rs$. In the \emph{max-exposure
	problem}, given an integer parameter $k$, we want to delete $k$ rectangles from $\rs$ so as 
	to maximize the number of exposed points.
	We show that the problem is $\np$-hard and assuming plausible complexity conjectures is
	also hard to approximate even when rectangles in $\rs$ are translates of two fixed rectangles. 
	However, if $\rs$ only consists of translates of a single rectangle, we present a polynomial-time
	approximation scheme. For range space defined by general rectangles, we present a simple $O(k)$ bicriteria 
	approximation algorithm; that is by deleting $O(k^2)$ rectangles, 
	we can expose at least $\Omega(1/k)$ of the optimal number of points.
\end{abstract}



\section{Introduction}
	Let $S = (P, \rs)$ be a geometric set system, also called a \emph{range space}, where $P$
	is a set of points and each $R \in \rs$ is a subset of $P$, also called a range.
	We are primarily interested in range spaces defined by a set of points in two dimensions and ranges 
	defined by axis-aligned rectangles. We say that a point $p \in P$ is \emph{exposed} if no range in $\rs$
	contains $p$. The \emph{max-exposure} problem is defined as follows: given a range space 
	$(P, \rs)$ and an integer parameter $k \geq 1$, remove $k$ ranges from $\rs$ so that
	a maximum number of points are exposed. That is, we want to find a subfamily $\rs^* \subseteq \rs$
	with $|\rs^*| = k$, so that the number of exposed points in the (reduced) range space $(P, \rs \setminus \rs^*)$
	is maximized.

	The max-exposure problem arises naturally in many geometric coverage settings. For instance,
	if points are the location of clients in the two-dimensional plane, and ranges correspond to 
	coverage areas of facilities, then exposed points are those not covered by any facility. The max-exposure 
	problem in this case gives a worst-case bound on the number of clients that can be exposed
	if an adversary disables $k$ facilities. Similarly, in distributed sensor networks, ranges
	correspond to \emph{sensing zones}, points correspond to physical assets being
	monitored by the network, and the max-exposure problem computes the number of assets
	exposed when $k$ sensors are compromised.

	More broadly, the max-exposure problem is related to the densest $k$-subgraph problem
	in \emph{hypergraphs}. In the \emph{densest $k$-subhypergraph} problem, we are given a hypergraph $H = (X, E)$,
	and we want to find a set of $k$ vertices with a maximum number of induced hyperedges. In general
	hypergraphs, finding $k$-densest subgraphs is known to be (conditionally) hard to approximate
	within a factor of $n^{1-\eps}$, where $n$ is the number of vertices. 
	The max-exposure problem is equivalent to the densest $k$-subhypergraph problem on a \emph{dual} hypergraph,
	whose vertices $X$ corresponds to the ranges $\rs$, and whose hyperedges correspond to the set of 
	points $P$. Specifically, each point $p \in P$ corresponds to a hyperedge adjacent to the set of ranges 
	containing the point $p$. In the rest of the paper, we will use $n = |\rs|$ for the number of ranges in $\rs$
	and $m = |P|$ for the number of points. We show that if the range space is defined by \emph{convex polygons},
	then the max-exposure problem is just as hard as the densest $k$-subhypergraph problem. However, for ranges 
	defined by \emph{axis-aligned rectangles}, one can achieve better approximation.
	In particular, we obtain the following results. 

	\begin{itemize}
		\item We show that the max-exposure problem is \np-hard and assuming the \emph{dense vs random} conjecture~\cite{chlamtavc2017minimizing},
	it is also hard to approximate better than a factor of $O(n^{1/4})$ even if the range space is defined by 
	\emph{only two types of rectangles} in the plane. For range space defined by convex polygons,
					we show that max-exposure is equivalent to densest $k$-subhypergraph problem, which is hard
					to approximate within a factor of $O(n^{1-\eps})$.

		\item When ranges are defined by translates of a \emph{single} rectangle, we give a polynomial-time approximation
					scheme (PTAS) for max-exposure. The PTAS stands in sharp contrast to the inapproximability of ranges 
					defined by \emph{two} types of rectangles. Moreover, as an easy consequence of this result, we obtain a 
					constant approximation when the ratio of longest and smallest side of rectangles in $\rs$ is bounded by a constant.
					However, we do not know if max-exposure with translates of a single rectangle can be solved in 
					polynomial time or is NP-hard.

		\item For ranges defined by arbitrary rectangles, we present a simple greedy algorithm that achieves
					a bicriteria $O(k)$-approximation. That is, if the optimal number of points exposed is $m^*$, the
					algorithm picks a subset of $k^2$ rectangles such that the number of points exposed is 
					at least $m^*/ck$, for some constant $c$.
					No such approximation is possible for general hypergraphs.
					If rectangles in $\rs$ have a bounded aspect ratio, the approximation improves to $O(\sqrt{k})$.
					For pseudodisks with \emph{bounded-ply} (no point in the plane is contained in more than a 
					constant number of ranges), this algorithm achieves a constant approximation.
	\end{itemize}

	The PTAS is obtained by first optimally solving a restricted max-exposure instance where all points 
	are contained in a unit square using dynamic programming in polynomial time.
	Next, we carefully combine them to obtain an optimal solution in $(nm)^{O(h^2)}$ time 
	for the case when input points lie in a $h \times h$ square. Applying well known shifting
	techniques on this gives us the PTAS. Both bicriteria algorithms are obtained by carefully
	assigning the points to ranges and applying greedy strategies.

	\subparagraph{Related Work}
	Coverage and exposure problems have been widely studied in geometry and graphs. In the classical
	\emph{set cover} problem, we want to select a subfamily of $k$ sets that cover the 
	maximum number of items (points)~\cite{feige1998threshold,fowler1981optimal}.
	For the set cover problem, the classical greedy algorithm achieves a factor $\log n$ approximation for the number of
	sets needed to cover all the items, or factor $(1- 1/e)$ approximation for the number of items covered by using
	exactly $k$ sets. Similarly, in geometry, the art gallery problems explore coverage of polygons using a minimum number of guards. 
	Unlike coverage problems where greedy algorithms deliver reasonably good approximation, the exposure
	problems turn out to be much harder. Specifically, choosing $k$ sets whose union is of \emph{minimum size} is much harder
	to approximate with a conditional inapproximability of $O(n^{1-\eps})$ where $n$ is the number of elements,
	or $O(m^{1/4-\eps})$ where $m$ is the number of sets~\cite{chlamtavc2017minimizing}. This so-called \emph{min-union}
	problem is essentially the complement of the densest $k$-subhypergraph problem on hypergraphs~\cite{chlamtac2018densest}. 
	The densest $k$-subgraph problem for graphs has a long history~\cite{feige2001dense, asahiro2000greedily, arora1999polynomial, bhaskara2010detecting}.
	The paper~\cite{chlamtac2018densest} also studies the special case of an \emph{interval hypergraph} $H = (V, E)$,
	whose vertices $V$ is a finite subset of $\mathbb{N}$ and for each edge $e \in E$ there are values
	$a_e, b_e \in \mathbb{N}$ such that $e = \{i \in V : a_e \leq i \leq b_e \}$. 
	That is, vertices are integer points and edges are intervals containing them. 
	They show that this restricted case can be solved in polynomial time. The corresponding max exposure
	instance is when ranges $\rs$ are intervals $R_i = (a_i, b_i)$ on the real line. As discussed
	later, this 1-D case can also be solved in polynomial time. Moreover, we show that good approximations
	can also be obtained for some geometric objects in two dimensions.
	
	The coverage problems have also been studied for geometric set systems where improved approximation bounds 
	are possible using the $VC$ dimension~\cite{agarwal2014near, bronnimann1995almost, mustafa2014settling}.
	Multi-cover variants, where each input point must be covered by more than one set, are 
	studied in~\cite{chekuri2012set,cygan2011approximation}.
	The geometric constraint removal problem~\cite{bandyapadhyay2018improved,eiben2018improved}, 
	where given a set of ranges, the goal is to \emph{expose} a path between two given points by deleting 
	at most $k$ ranges (a path is exposed if it lies in the exterior of all ranges),
	is also closely related to the max-exposure problem.
 	Even for simple shapes such as unit disks (or unit squares)~\cite{BeregK09, KormanLSS18}, no PTAS is known for this problem.

	The remainder of the paper is organized as follows. In Section~\ref{sec:hardness}, we discuss our hardness results
	followed by the bicriteria $O(k)$-approximation in Section~\ref{sec:greedy}. In Section~\ref{sec:ptas}, we
	study the case when $\rs$ consists of translates of a fixed rectangle and describe a PTAS for it.
	Finally, in Section~\ref{sec:extensions}, we use these ideas to obtain a bicriteria $O(\sqrt{k})$-approximation
	when the aspect ratio of rectangles in $\rs$ is bounded by a constant.

\section{Hardness of Max-Exposure}
	\label{sec:hardness}
	We show that the max-exposure problem for geometric ranges is both \np-hard, and inapproximable.
	We begin by reducing the densest $k$-subgraph on bipartite graphs (\emph{bipartite-DkS}) to the max-exposure 
	problem; the known \np-hardness of biparite-DkS then implies the hardness for max-exposure. Moreover, we show
	that bipartite-DkS is hard to approximate assuming the \emph{dense vs random} conjecture, thereby establishing
	the inapproximability of max-exposure.

	In the \emph{bipartite-DkS} problem, we are given a  bipartite graph $G = (A, B, E)$, an integer $k$, 
	and we want to compute a set of $k$ vertices 
	such that the induced subgraph on those $k$ vertices has the maximum number of edges.
	Given an instance $G=(A, B, E)$ of bipartite-DkS, we construct a max-exposure instance as follows.

\begin{figure}[t!]
		\begin{minipage}[t]{0.40\linewidth}
		\centering
		\includegraphics{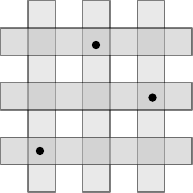}
		\caption{Reducing bipartite-DkS to max-exposure with axis-aligned rectangles.}
		\label{fig:nphard-rect}
		\end{minipage}
		\hfill
		\begin{minipage}[t]{0.55\linewidth}
		\centering
		\includegraphics{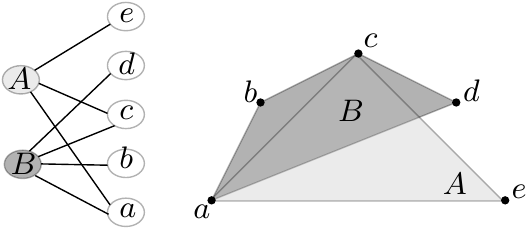}
		\caption{Reducing densest $k$-subhypergraph problem to max-exposure. Hypergraph vertices $A, B$ shown as convex ranges.}
		\label{fig:convexHard}
		\end{minipage}
	\end{figure}

	Let $R_1 = [0, \eps] \times [0, n]$ be a thin vertical rectangle and $R_2 = [0, n] \times [0, \eps]$
	be a thin horizontal rectangle. For each vertex $v_i \in A$, we create a copy $R_i$ of $R_1$, and
	place it such that its lower-left corner is at $(i, 0)$. Similarly, for each vertex $v_j \in B$, 
	we create a copy $R_j$ of $R_2$, and place it such that its lower-left corner is at $(0, j)$. 
	These $|A| + |B|$ rectangles create a checkerboard arrangement, with $|A| \times |B|$ cells of intersection.
	For each edge $(v_i, v_j) \in E$, we place a single point in the cell corresponding to intersection of $R_i$
	and $R_j$. It is now easy to see that $G$ has a $k$-subgraph with $m^*$ edges if and only if we can 
	expose $m^*$ points in this instance by removing $k$ rectangles: the removed rectangles are 
	exactly the $k$ vertices chosen in the graph, and each exposed point corresponds to the 
	edge included in the output subgraph. (See also Figure~\ref{fig:nphard-rect}.)
	We will later make use of this reduction, and therefore state it as the following lemma.

	\begin{lemma}
		\label{lemma:bipartiteDkS}
		The max-exposure problem is at least as hard as bipartite-DkS.
	\end{lemma}
	Since bipartite-DkS is known to be \np-hard~\cite{FeigeDensestSubgraph}, we have the following.
	\begin{theorem}
		Max-exposure problem with axis-aligned rectangles is \np-hard.
	\end{theorem}
	
	\subsection{Hardness of Approximation} 
	The construction in the preceding proof shows that max-exposure with rectangles is at least
	as hard as bipartite-DkS problem. Moreover, the geometric construction uses translates
	of \emph{only two rectangles} $R_1, R_2$. In the following, we show that even with such a restricted range space,
	the problem is also hard to approximate. To that end we prove that bipartite-DkS cannot be approximated
	better than a factor $O(n^{1/4})$, where $n$ is the number of vertices in this graph.
	More precisely, if the densest subgraph over $k$ vertices has $m^*$ edges, 
	it is hard to find a subgraph over $k$ vertices that contains $\Omega(m^*/n^{\frac{1}{4} - \eps})$ edges 
	in polynomial time.  This hardness of approximation is conditioned on the so-called \emph{dense vs random} 
	conjecture~\cite{chlamtavc2017minimizing} stated as follows.
	
	Given a graph $G$, constants $0 < \alpha, \beta < 1$, 
	and a parameter $k$, we want to distinguish between the following two cases.
		\begin{enumerate}
			\item (\textsc{Random})~ $G = G(n, p)$ where $p = n^{\alpha-1}$, that is, $G$ has average degree approximately $n^\alpha$.
			\item (\textsc{Dense})~ $G$ is adversarially chosen so that the densest $k$-subgraph of $G$ has average degree $k^\beta$.
		\end{enumerate}
	The conjecture states that for all $0 < \alpha < 1$, sufficiently small $\eps > 0$, and for all $k \leq \sqrt{n}$,
	one cannot distinguish between the \emph{dense} and \emph{random} cases in polynomial time (w.h.p), 
	when $\beta \leq \alpha - \eps$.
	
	In order to obtain hardness guarantees using the above conjecture, one needs to find the
	`distinguishing ratio' $r$, that is the least multiplicative gap between the optimum solution
	for the problem on the dense and random instances. If there exists an algorithm with an approximation 
	factor significantly smaller than $r$, then we would be able to use it to distinguish between the
	dense and random instances, thereby refuting the conjecture.
	We obtain the following result for densest $k$-subgraph problem on bipartite graphs.
	\begin{lemma}
		\label{lemma:bipartiteDksHard}
		Assuming that \emph{dense vs random} conjecture is true, the densest $k$-subgraph problem on bipartite
		graphs is hard to approximate better than a factor $O(n^{1/4})$ of optimum.
	\end{lemma}
	
	\begin{proof}
				Let $G'=(V',E')$ be a graph sampled either from the dense or from the random instances.
				We construct a bipartite graph $G = (A, B, E)$ as follows. 
				For every vertex $v \in V'$, add a vertex $v_a$ to $A$ and $v_b$ to $B$. 
				For every edge $e = (u, v) \in E'$, we add the pair of edges $e_1 = (u_a, v_b)$ and $e_2 = (v_a, u_b)$ to $E$. 
				That is, every edge $e \in E'$ is mapped to two copies 
				$e_1, e_2 \in E$ and we define $e$ to be their \emph{parent} edge as 
				$\textit{par}(e_1) = \textit{par}(e_2) = e$. 
				Similarly, for a vertex $u \in V'$ and its two copies $u_a, u_b \in V$,
				we define $\textit{par}(u_a) = \textit{par}(u_b) = u$.
				We say that $G$ is \emph{dense} if the underlying graph $G'$ was sampled from the dense case,
				otherwise we say that $G$ is \emph{random}.

				Consider a set of $k^*= 2k$ vertices in $G$. If $G$ came from the dense case, there must be a set
				of $2k$ vertices that have $2k^{\beta+1}$ edges between them. So the number of edges in 
				dense case $m^*_d \geq 2k^{\beta+1}$.
				Otherwise, we are in the random case. Consider the optimal set of $k^* = 2k$ vertices $V^*$
				that maximizes the set $E^*$ of edges in the induced subgraph $G[V^*]$. 
				Now consider the corresponding set of vertices $V_p = \{\textit{par(v)} ~|~ v \in V^* \}$ of the
				original graph $G'$ and the set of edges $E_p$ in the induced subgraph $G'[V_p])$.
				We have that $|V_p| \leq |V^*| = 2k$ and $|E_p| \geq |E^*|/2$ because for each
				edge $e = (u, v) \in E^*$, we will have the edge $par(e) = (\textit{par}(u), \textit{par}(v)) \in E_p$.
				Since $|V_p| \leq 2k$ and we are in the random case, we can upperbound the number of edges in $E_p$ as
				the number of edges in the densest subgraph of $G(n, n^{\alpha-1})$ over $2k$ vertices. 
				This is $\tilde{O}(\max(2k, 4k^2n^{\alpha-1}))$ w.h.p. where $\tilde{O}$ ignores logarithmic factors.
				Therefore the optimum number of edges in the random case 
				is $m^*_r = |E^*| \leq 2|E_p| = \tilde{O}(\max(k, k^2n^{\alpha-1}))$ w.h.p.

				Choosing $k = n^{1/2}$, $\alpha = \frac{1}{2}$, $\beta = \frac{1}{2} - \eps$, gives us $m^*_r = \tilde{O}(n^{1/2})$ w.h.p.
				and $m^*_d = \tilde{\Omega}(n^{\frac{3-2\eps}{4}})$. If we could approximate this problem within a
				factor $O(n^{1/4 - \eps})$, then in the dense case, the number of edges computed by this 
				approximation algorithm is $\tilde{\Omega}(n^{\frac{1+ \eps}{2}})$ which is strictly more than the maximum possible edges
				in the random case. Therefore, we would be able to distinguish between dense and random cases,
				and thereby refute the conjecture for these values of $\alpha, \beta$ and $k$. 
	\end{proof}
	
	Using the same construction as in Lemma~\ref{lemma:bipartiteDkS}, we obtain the following.
	\begin{corollary}
		Assuming the dense vs random conjecture, max-exposure with axis-aligned rectangles is 
		hard to approximate better than factor $O(n^{1/4})$ of optimum.
	\end{corollary}

\subsection{Hardness of Max-exposure with Convex Polygons} 
We now show that the max-exposure problem is \emph{equivalent} to the densest $k$-subhypergraph problem for 
\emph{general hypergraphs} when the range space $(P, \rs)$ is defined by convex polygons.
In one direction, the max-exposure instance $(P, \rs)$ naturally corresponds to a hypergraph $H = (\rs, P)$ whose 
vertices are the ranges and the edges correspond to points and are defined by the containment relationship. Clearly, 
the densest $k$-subhypergraph corresponds to the set of $k$ ranges deleting which exposes maximum number of points.
For the other direction, we have the following lemma. (See also Figure~\ref{fig:convexHard}.)

\begin{lemma}
	Given a hypergraph $H=(X, E)$, one can construct a max-exposure instance with convex ranges $\rs$
	and points $P$ such that the densest $k$-subhypergraph of $H$ corresponds to a solution of max-exposure.
\end{lemma}
\begin{proof}
For each edge $e \in E$ of the hypergraph, add a point $p_e \in P$. 
We place all the points of $P$ in convex position. Let $v \in X$ be a vertex and $E_v$ 
be the set of hyperedges adjacent to $v$. 
Since points in $P$ are in convex position, any subset of $P$ forms a convex polygon. 
Therefore, for every $ v \in V$, we can draw a convex polygon $R_v \in \rs$ whose corners 
are the point set corresponding to the hyperedges $E_v$.
The polygons will likely overlap in the convex region but for every point $p_e \in P$, the polygons containing $p_e$
are precisely the ones that have $p_e$ as its corner. Therefore, $p_e$ is exposed if and only if all
vertices of the hyperedge $e$ are selected.
\end{proof}

\section{\boldmath A Bicriteria $O(k)$-approximation Algorithm}
\label{sec:greedy}
In this section, we present a simple approximation algorithm for the max-exposure problem
that achieves bicriteria $O(k)$-approximation for range spaces defined by arbitrary
axis-aligned rectangles. Specifically, if the optimal number of points exposed is $m^*$, the
algorithm picks a subset of $k^2$ rectangles such that the
number of points exposed is at least $m^*/ck$, for some constant $c$.
In fact, the results hold for any polygonal range with $O(1)$ complexity.

This bicriteria approximation should be contrasted with the fact that no 
such approximation is possible for the densest $k$-subhypergraph problem: 
that is, one cannot compute a set of $O(k^b)$ vertices for any constant $b$
such that the number of edges in the induced subhypergraph is at least optimal.
Thus the geometric properties of the range space have a significant impact on 
the problem complexity.
In particular, if $\rs$ consists of rectangle ranges, we show that the
following strategy picks a subset of $\alpha k$ ranges such that 
the number of points exposed is at least $\alpha m^* / (ck^2)$, 
for a parameter $1 \leq \alpha \leq k$ and constant $c$ that will be fixed later.
Choosing $\alpha = k$ gives us the claimed bound.

Our algorithm is essentially greedy. We divide the points into maximal equivalence
classes, where each class is the maximal subset of points belonging to the same subset
of ranges. We define $\rs(p)$ as the set of ranges that contain a point $p \in P$,
and remove all points that are contained in more than $k$ ranges, since they can
be never exposed in the optimal solution. Therefore, without loss of generality,
we can assume that $|\rs(p)| \leq k$ for all points $p \in P$. The rest of the algorithms
is as follows.

\begin{algorithm}
\begin{enumerate}
	\item Partition $P$ into a set $\cg$ of groups  where each group $G_i \in \cg$ is 
        an equivalence class of points that are contained in the same set of ranges.
				That is, for any $p \in G_i, ~p' \in G_j$, we have $\rs(p) = \rs(p')$ if $i = j$
				and $\rs(p) \neq \rs(p')$, otherwise.

	\item Sort the groups in $\cg$ by decreasing order of their size $|G_i|$ and
				select the ranges in first $\alpha$ groups for deletion. 
				
	\item Return $m' = \sum_{1 \leq i \leq \alpha} |G_i|$ as the number of points exposed.
\end{enumerate}
 \caption{Greedy-Bicriteria}
 \label{alg:greedy-bicriteria}
\end{algorithm}

In Algorithm~\ref{alg:greedy-bicriteria}, observe that every point in the $i$th group $G_i$ 
is contained in the same set of ranges, which we denote by $\rs(G_i)$. Moreover, we have $|\rs(G_i)| \leq k$. 
Therefore, the total number of ranges that we delete in Step~2 is at most $\alpha k$.
It remains to show that the number of points exposed $m'$ is at least $\alpha m^* / ck^2$.
\begin{lemma}
	\label{lemma:bicriteria}
	Let  $m'$ be the number of points exposed by the algorithm \emph{Greedy-Bicriteria},
	and let $m^*$ be the optimal number of exposed points,
	Then, $m' \geq \alpha m^* / ck^2$.
\end{lemma}
\begin{proof}
	Consider the optimal set $\rs^*$ of $k$ ranges that are deleted, and let $P^*$ be the set of exposed points.
	We partition the set of points $P^*$ into groups $\cg^*$ as before, 
	such that each group $G_i^* \in \cg^*$ is identified by the range set $\rs(G_i^*) = \rs(p)$, 
	for any $p \in G_i^*$. Since $P^* \subseteq P$, we must have that $\cg^* \subseteq \cg$.
	This holds because for every group $G^*_i \in \cg^*$ there must be a group $G_j \in \cg$ such that
	$\rs(G_i^*) = \rs(G_j)$. Moreover since $P^*$ is the maximum set of points that can be exposed,
	we must have that $G_i^* = G_j$. Finally, we note that the number of groups $|\cg^*|$ is bounded
	by the number of cells in the arrangement of ranges in $\rs^*$ which is at most $ck^2$
	for some fixed constant $c$, for all $O(1)$-complexity ranges.
	If the groups in $\cg$ are arranged by decreasing order of their sizes, we have that

	\begin{align*}
		m^*   ~&=~ \sum_{1 \leq i \leq |\cg^*|} |G_i^*| 
				  ~\leq~ \sum_{1 \leq i \leq |\cg^*|} |G_i| 
				  ~\leq~ \sum_{1 \leq i \leq ck^2} |G_i| 
	\end{align*}
	\begin{align*}
				  ~&\leq~ \frac{ck^2}{\alpha} \sum_{1 \leq i \leq \alpha} |G_i| 
					~=~ \frac{ck^2}{\alpha} \cdot m'
	\end{align*}
\end{proof}

The parameter $\alpha$ can be tuned to improve the approximation guarantee with respect
to one criterion (say the number of exposed points) at the cost of other.
With $\alpha = k$, the algorithm exposes at least $\Omega(m^*/k)$ by removing $k^2$ ranges.
As for the running time, a simple implementation of the algorithm can be made to
run in $O(mn \log m)$ time: we can build the point-range containment relation in $O(mn)$
time, partitioning the point set into groups takes an additional $O(mn \log m)$ time.

\subsection{Constant Approximation for Pseudodisks with Bounded-ply} 
If the range space $\rs$ consists of pseudodisks 
of \emph{bounded-ply} (no point in the plane is contained in more than a constant number $\rho$ pseudodisks), 
then the algorithm Greedy-Bicriteria achieves a constant approximation. Due to the bounded-ply restriction,
we have that the number of pseudodisks containing the points of group $G_i$ is $|\rs(G_i)| \leq \rho$, and
therefore number of pseudodisks that are removed in Step 2 of the algorithm is also at most $\alpha \rho$.
Moreover, the number of cells in an arrangement of $k$ pseudodisks with depth at most $\rho$ is $O(\rho k)$~\cite{ClarksonS89}.
Therefore, we can bound the number of groups of the optimal solution $|\cg^*|$ in the proof for Lemma~\ref{lemma:bicriteria} to be
at most $c\rho k$. This gives us that the number of points exposed $m' \geq \alpha m^* / c\rho k$, where $m^*$ is the number
of points exposed by the optimal solution.
\begin{lemma}
	If the range space $\rs$ consists of pseudodisks of \emph{bounded-ply} $\rho$, then algorithm \emph{Greedy-Bicriteria}
	exposes at least $\alpha m^* / c\rho k$ points by deleting at most $\alpha \rho$ pseudodisks, 
	where $ 1 \leq \alpha \leq k$.
\end{lemma}

Choosing $\alpha = k$, the algorithm achieves a bicriteria $O(\rho)$-approximation.
With $\alpha = k/\rho$, the algorithm exposes at least $1/c\rho^2$ fraction of the optimal number of points
by deleting $k$ ranges.

\section{A PTAS for Unit Square Ranges}
\label{sec:ptas}
We have seen that max-exposure is hard to approximate even if the ranges are translates of
two types of rectangles. We now describe an approximation
scheme when the ranges are translates of a \emph{single} rectangle. In this case, we can
scale the axes so that the rectangle becomes a \emph{unit square} without changing
any point-rectangle containment. Therefore, we can assume that our ranges are all unit squares.
The problem is non-trivial even for unit square ranges, and as a warmup we first solve the
following special case: \emph{all the points lie inside a unit square.}
We develop a dynamic programming algorithm to solve this case exactly, and then use
it to design an approximation for the general set of points.

\subsection{Exact Solution in a Unit Square}
\label{sec:dp-unit-square}
		
\begin{figure}[h!]
		\begin{minipage}[t]{0.50\linewidth}
		\centering
			\includegraphics{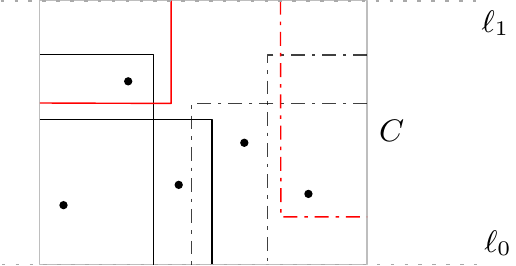}
			\caption{Max-exposure in a unit square $C$.  
						Type 0 ranges are drawn with solid lines, Type 1 
						ranges are dash-dotted.}
			\label{fig:anchored}
		\end{minipage}
		\hfill
		\begin{minipage}[t]{0.45\linewidth}
		\centering
		\includegraphics{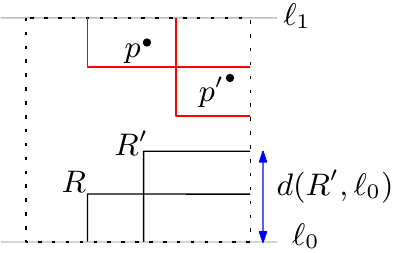}
		\caption{An example of \emph{closer} relationship. Point $p$ is \closer to $\ell_1$ than $p'$. $R$ is closer to $\ell_0$ than $R'$.} 
		\label{fig:ordering}
	\end{minipage}
\end{figure}

We are given a max-exposure instance consisting of unit square ranges $\rs$ and a set of points $P$
in a unit square $C$. Without loss of generality, we can assume that the 
lower left corner of $C$ lies at origin $(0, 0)$ and all ranges in $\rs$ intersect $C$. 
We classify the ranges in $\rs$ to be one of the two types: (See also Figure~\ref{fig:anchored}). 

\begin{enumerate}
	\item[~]\textbf{\tz:} Unit square ranges that intersect $x = 0$.
	\item[~]\textbf{\to:} Unit square ranges that intersect $x = 1$.
\end{enumerate}

(A unit square range coincident with both $x=0$ and $x=1$ is assumed to be \tz). 
We draw two parallel horizontal lines $\ell_0 : y = 0$ and $\ell_1 : y =1$ 
coincident with bottom and top horizontal sides of $C$ respectively. 
We say that a range $R \in \rs$ is \emph{anchored} to  a line $\ell$ 
if it intersects $\ell$. Note that every $R \in \rs$ is anchored to exactly 
one of $\ell_0$ or $\ell_1$. (When $R$ is coincident with 
both $\ell_0$ and $\ell_1$, we say that it is anchored to $\ell_0$).

Moreover, for the rest of our discussion, let $x=x_i$ be a vertical line and define
$P_i \subseteq P$ to be the set of points that have $x$-coordinate at least $x_i$. 
In other words, $P_i$ is the set of \emph{ active points} at $x = x_i$.
Similarly, define $\rs_i \subseteq \rs$ to be the set of \emph{active ranges} that have at 
least one corner to the right of $x=x_i$. That is, $R \in \rs_i$ either intersects $x=x_i$ or 
lies completely to the right of it.

In order to gain some intuition, we will first consider the following two natural dynamic programming 
formulations for the problem. 

\subparagraph{DP-template-0}
Suppose that the points in $P$ are ordered by their increasing $x$-coordinates and
let $x_i$ be the $x$-coordinate of the $i$th point $p_i$.
We define a subproblem as $S(i, k', \rs_d)$ which represents the maximum 
number of points in $P_i$ that can be exposed by removing $k'$ ranges
from the range set $\rs_i \setminus \rs_d$.
If we define $x_0 = 0$, then $S(0, k, \emptyset)$ gives the optimal number of exposed points for
our problem.

Let $k_i = |\rs(p_i) \setminus \rs_d|$ be the number of new ranges in $\rs_i$ that contain $p_i$.
Then, we can can express the subproblems at $i$ in terms of subproblems at $i+1$ as follows.
\begin{align*}
	S(i,~ k',\rs_d)  = \max
	\begin{cases}
		S(i+1,~ k' - k_i, ~\rs_d \cup \rs(p_i)) ~+~ 1          &~\hspace{2em}\text{expose $p_i$}\\
		S\left(i+1,~ k', ~\rs_d\right) 										     &~\hspace{2em}\text{$p_i$ not exposed}
	\end{cases}
\end{align*}

Roughly speaking, at $x=x_i$ which is the \emph{event} corresponding to a point $p_i \in P$,
we have two choices :  \emph{expose $p_i$} or \emph{do not expose $p_i$}. 
If we expose $p_i$, we pay for deleting the ranges in $\rs_i \setminus \rs_d$ that contain $p_i$ and 
mark them as deleted by adding to the \emph{deleted range set} $\rs_d$. 
It is easy to see that this correctly computes the 
optimal number of exposed points since we charge for every deletion exactly once.
However, there is one complication: a priori it is not clear how 
to bound the number of range subset $\rs_d$ used by this dynamic program. 
We later argue that the geometry of
range space for \tz ranges allows us to use only a polynomial number of choices.

\subparagraph{DP-template-1}
An alternative approach is to consider both \emph{point} and \emph{begin-range} events.
That is, $x=x_i$ is either incident to a point $p_i \in P$ or to the left vertical side
of a range $R_i \in \rs$. 
Then, we can define a subproblem by the tuple $S(i, k', P_f)$
which represents the maximum number of points in $(P_i \setminus P_f)$ that can be exposed
by removing $k'$ ranges in $\rs_i$.
If we define $x_0 = 0$, then $S(0, k, \emptyset)$ gives the optimal number of exposed points. 
Let $P(R_i) \subseteq P$ be the set of points contained in the range $R_i$,
then we have the following recurrence.

\begin{align*}
	S(i,~ k',P_f)  ~&=~ \max
	\begin{cases}
		S(i+1,~ k'-1,~P_f) 						    						&\hspace{2em}\text{delete range $R_i$}\\
		S(i+1,~ k',~P_f \cup P(R_i)) 						      &\hspace{2em}\text{$R_i$ not deleted}\\
	\end{cases}
	\\
	&\text{\hspace{2em}\emph{(event $x=x_i$ was beginning of a range $R_i \in \rs_i$)}}\\ 
	~&=~ \max
	 \begin{cases}
			S(i+1,~ k',P_f) 	   						  &\hspace{2em}\text{if $p_i \in P_f$, cannot expose $p_i$} \\
			S(i+1,~ k',P_f) + 1  		        &\hspace{2em}\text{otherwise, expose $p_i$} \\
	 \end{cases}
	\\
	&\text{\hspace{2em}\emph{(otherwise, event $x=x_i$ was a point $p_i \in P_i$)}}
\end{align*}
In the above formulation, at each \emph{begin-range} event for some $R_i \in \rs_i$, 
we have two choices: \emph{delete $R_i$} or \emph{do not delete $R_i$}.
If $R_i$ was deleted, we reduce the budget $k'$ by one. Otherwise, if $R_i$ was
not deleted, we can never expose the points in $P(R_i)$, and therefore we 
add $P(R_i)$ to the \emph{forbidden point set} $P_f$.
The correctness of the dynamic program follows from the fact that for every point $p_i$,
all the ranges containing it must begin before $x=x_i$, and we expose $p_i$ only if
those ranges were \emph{deleted}. 
Again, it is not obvious how many different subsets $P_f$ are needed by the
dynamic program. However, we will later show that by keeping
track of polynomial number of sets $P_f$, we can solve max-exposure with \to ranges.

We note that the \tz and \to ranges may superficially seem symmetric but once we 
fix the order of computing subproblems, they become structurally different. 
Therefore, we would need slightly different techniques to handle each type. 
For the ease of exposition, we present dynamic programs for \tz and \to ranges separately
and finally combine them. Also note that if the ranges in $\rs$ are intervals on the 
real line (max exposure in 1D), then both DP-template-0 and DP-template-1 can be 
easily applied to obtain a polynomial time algorithm.

We will now define the following ordering relations that will be useful later.
Let $\ell$ be a horizontal line, and let $d(p, \ell)$ denote the orthogonal distance of $p \in P$ from $\ell$.
If $p, p' \in P$ are two points, we say that $p$ is \closer to $\ell$ than $p'$ if $d(p, \ell) < d(p', \ell)$.
Similarly, for a range $R \in \rs$ that is anchored to $\ell$,  let $d(R, \ell)$ be the vertical
distance \emph{inside the unit square $C$} between $\ell$ and the side of $R$ parallel to $\ell$.
If $R, R' \in \rs$ are two ranges, we say that $R$ is \closer (or equivalently $R'$ is \farther) from $\ell$ 
if both $R, R'$ are anchored to $\ell$ and $d(R, \ell) < d(R', \ell)$. (See Figure~\ref{fig:ordering}.)

\subsubsection{Max-exposure with \tz Ranges}
	Recall that $\tz$ ranges intersect the vertical lines $x=0$ and are anchored to either
	$\ell_0$ or $\ell_1$. We will apply the formulation discussed
	in \emph{DP-template-0}. The key challenge here is to bound the number of possible deleted range sets $\rs_d$.
	Towards that end, we make the following claim. Recall that $\rs_i$ is the set of active ranges at $x=x_i$.
	\begin{lemma}
		Let $q_0, q_1$ be the two exposed points strictly to the left of $x=x_i$ that are
		closest to $\ell_0$ and $\ell_1$ respectively. Then our dynamic program only needs
		to consider the set of deleted ranges $\rs_d = \rs_i \cap (\rs(q_0) \cup \rs(q_1))$ 
		at $x=x_i$ conditioned on $q_0, q_1$. 
	\end{lemma}
	\begin{proof}
		Observe that since $\rs$ consists of \tz ranges, every range in $\rs_i$ must 
		intersect the vertical line $x=x_i$. Suppose we partition $\rs_i$ into ranges 
		$\rs^0_i$ that are anchored to $\ell_0$ and $\rs^1_i$ that are anchored to $\ell_1$.
		Let $P' \subseteq P$ be the set of all \emph{exposed points} strictly to the left of $x=x_i$.
		Observe that for all $p \in P'$, any range $R \in \rs^0_i$ that contains $p$ must also contain $q_0$.
		Therefore, we must have $\rs^0_i \cap \rs(p) \subseteq \rs^0_i \cap \rs(q_0)$, for all $p \in P'$.
		Similarly,  $\rs^1_i \cap \rs(p) \subseteq \rs^1_i \cap \rs(q_1)$, for all $p \in P'$.
		This gives us $\bigcup_{p\in P'} \rs_i \cap \rs(p) = \rs_i \cap (\rs(q_0) \cup \rs(q_1))$.
		Therefore, the set $\rs_d$ consists of all the active ranges that contain at least one exposed point
		and were therefore deleted to the left of $x=x_i$.
	\end{proof}

	Therefore, if our dynamic program remembers the exposed points $q_0, q_1$, then
	we can compute the deleted range set $\rs_d = \rs_i \cap (\rs(q_0) \cup \rs(q_1))$ at $x=x_i$. 
	There are $O(m^2)$ choices for the pair $q_0, q_1$, 
	so the number of possible sets $\rs_d$ is also $O(m^2)$. 
	We can therefore identify our subproblems by the tuple $S(i,~ k',~ q_0,~ q_1)$ which 
	represents the maximum number of exposed points with $x$-coordinates $x_i$ or higher
	using $k'$ rectangles from the set $\rs_i \setminus \rs_d$.
	With $k_i = |\rs(p_i) \setminus \rs_d|$, we obtain the following recurrence:
\begin{align*}
	S&(i,~ k',~ q_0,~ q_1) ~=~\\
	&\max
	\begin{cases}
		S\left(i+1,~ k' - k_i,~\textit{closer}(q_0, p_i), ~\textit{closer}(q_1, p_i) \right) ~+~ 1  &\hspace{1em}\text{expose $p_i$}\\
		S\left(i+1,~ k',~ q_0,~ q_1\right) 																				&\hspace{1em}\text{$p_i$ not exposed}
	\end{cases}
\end{align*}
	where the function $\textit{closer}(q_0, p_i)$ returns whichever of $q_0, p_i$ is closer to $\ell_0$,
	and $\textit{closer}(q_1, p_i)$ returns whichever of $q_1, p_i$ is closer to $\ell_1$.
	The optimal solution is given by $S(0,~ k,~ \pbf,~\ptf)$, where $\pbf = (0, 1)$ and $\ptf = (0, 0)$ are two artificial 
	points with $\rs(\pbf) = \rs(\ptf) = \emptyset$ (not contained in any range).
	The base case is defined by the rightmost event at vertical line $x = 1$ and 
	is initialized with zeroes for all $q_0, q_1$ and $k' \geq 0$. 
	Any subproblem with $k' < 0$ has value $-\infty$.

\subsubsection{Max-exposure with \to Ranges}
Next we consider the case when we only have \to ranges in $\rs$.
Unfortunately in this case, our previous dynamic program does not work
and we need to remember a different set of parameters.
More precisely, we will apply the formulation discussed in \emph{DP-template-1},
and bound the number of possible forbidden point sets $P_f$. Recall that
$P_i$ is the set of active points at $x=x_i$ (with $x$-coordinate $x_i$ or higher).

\begin{figure}[h!]
		\begin{minipage}[t]{0.40\linewidth}
		\centering
			\includegraphics{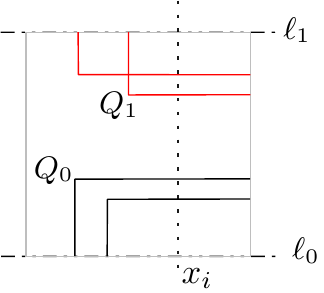}
			\caption{Undeleted ranges $Q_0$ and $Q_1$ farthest from $\ell_0$ and $\ell_1$ respectively.} 
			\label{fig:rememberRanges}
		\end{minipage}
		\hfill
		\begin{minipage}[t]{0.55\linewidth}
		\centering
			\includegraphics{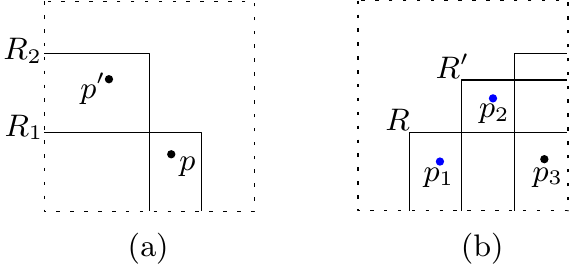}
			\caption{Remembering one of $R_1, R_2$ in (a) or one of $p_1, p_2$ in (b) is not sufficient.}
			\label{fig:anchoredCases}
		\end{minipage}
\end{figure}

\begin{lemma}
		Let $Q_0, Q_1$ be two ranges that begin to the left of $x=x_i$ and were \emph{not deleted}.
		Moreover, $Q_0$ is anchored to and is farthest from $\ell_0$.
		Similarly  $Q_1$ is anchored to and is farthest from $\ell_1$
		(Figure~\ref{fig:rememberRanges}).
		Then the forbidden point set at $x=x_i$ is given by 
		$P_f = P_i \cap (P(Q_0) \cup P(Q_1))$, where $P(Q)$ is the set of points contained in range $Q$.
\end{lemma}
	
\begin{proof}
		Recall that the set $\rs_i$ consists of ranges that have at least one corner to the right
		of the vertical line $x=x_i$. Since we are dealing with \to
		ranges, every range that begins to the left of $x=x_i$ lies in $\rs_i$.
		Now let $\rs' \subseteq \rs_i$  be the set of ranges that begin to the left of $x=x_i$
		and were \emph{not deleted}.
		Here $P_i$ is the set of points in $P$ that have $x$-coordinate $x_i$ or higher.
		Now consider any range $R \in \rs'$. Recall that $R$ must be anchored to either $\ell_0$ or $\ell_1$.
		If $R$ was anchored to $\ell_0$, then every point of $P_i$ that lies in $R$ also lies in $Q_0$. 
		Otherwise $R$ was anchored to $\ell_1$, so every point of $P_i$ that lies in $R$ also lies in $Q_1$. 
		Therefore, $\bigcup_{R \in \rs'} ~(P_i \cap P(R)) = P_i \cap (P(Q_0) \cup P(Q_1))$,
		which is precisely the forbidden point set $P_f$.
	\end{proof}

Therefore, if our dynamic program remembers the ranges $Q_0$ and $Q_1$, we can compute 
the forbidden point set $P_f = P_i \cap (P(Q_0) \cup P(Q_1))$ at $x=x_i$. Since there are $O(n^2)$ choices
for the pair $Q_0, Q_1$, the number of possible sets $P_f$ is also $O(n^2)$.
We can now identify the subproblems by the tuple $S(i,~ k',~ Q_0,~ Q_1)$ 
which represents the maximum number of points in $P_i \setminus P_f$
that are exposed by deleting $k'$ ranges that begin on or after $x = x_i$. 
This gives us the following recurrence.

\begin{align*}
	&\hspace{-1em}S(i,~ k',~Q_0, ~Q_1)  ~=~ \\
	&\hspace{-1.5em}\max 
	  \begin{cases}
		S(i+1,~ k'-1,~Q_0, ~Q_1) 									    						&\text{delete range $R_i$}\\
		S(i+1,~ k',~\textit{farther}(Q_0, R_i),~\textit{farther}(Q_1, R_i)) &\text{$R_i$ not deleted}\\
	\end{cases}
	\\
	&\text{\hspace{2em}(\emph{event $x=x_i$ was beginning of a range $R_i \in \rs$})} \\
	~&\hspace{-1.5em}\max
	 \begin{cases}
			S(i+1,~ k',~Q_0, ~Q_1) 	   						  &\hspace{4em}\text{if $p_i \in P_f$, cannot expose $p_i$} \\
			S(i+1,~ k',~Q_0,~Q_1) + 1  		          &\hspace{4em}\text{otherwise, expose $p_i$} \\
	 \end{cases}
	\\
	&\text{\hspace{2em}\emph{(otherwise, event $x=x_i$ was a point $p_i \in P$)}}
\end{align*}
Here, the function \emph{farther} simply updates the ranges $Q_0, Q_1$ with $R_i$ if needed.
More precisely, if $R_i$ is anchored to $\ell_0$ and is farther from $\ell_0$ than $Q_0$, 
then $\textit{farther}(Q_0, R_i)$ returns $R_i$, otherwise it returns $Q_0$.
Similarly, if $R_i$ is anchored to $\ell_1$, and is farther from $\ell_1$ than $Q_1$, then
$\textit{farther}(Q_1, R_i)$ returns $R_i$, otherwise it returns $Q_1$.

The optimal solution is given by $P(0, k, \rbf, \rtf)$, where $\rbf, \rtf$ are two artificial
ranges of zero-width :  $\rbf$ is anchored to $\ell_0$ and is defined by corners $(0, 0)$ and  $(0, 1)$;
similarly, $\rtf$ is anchored to $\ell_1$ and is defined by corners $(0, 1)$ and $(1, 1)$.
\begin{remark}
\textnormal{
We note that remembering a constant number of exposed points $q_0, q_1$ (DP-template-0) or a 
constant number of undeleted ranges $Q_1, Q_2$ (DP-template-1) by themselves cannot solve \emph{both} \tz and \to ranges. 
For instance, in Figure~\ref{fig:anchoredCases}(a) with \tz ranges, if $R_1, R_2$ were both \emph{not deleted} but we remembered one of them, 
then we will incorrectly expose one of $p, p'$. Similarly in Figure~\ref{fig:anchoredCases}(b) with \to ranges,
if $p_1, p_2$ were both exposed but we only remembered one of them, we will pay for one of the ranges $R, R'$ again when we expose $p_3$.
However, since both the dynamic programs for \tz and \to ranges express subproblems at event $i$ in terms of
subproblems at event $i+1$, we can easily combine them with minor adjustments. }
\end{remark}
\begin{table}
\begin{center}
\begin{tabular}{ |l|l| } 
\hline
Notation  & Explanation \\ 
\hline
$\rs(p)$ & ranges containing point $p$\\ 
$P(R)$  & points contained in range $R$\\ 
$q_0, q_1$  & exposed points closest to $\ell_0, \ell_1$\\ 
$Q_0, Q_1$  & undeleted ranges farthest from $\ell_0, \ell_1$\\
$P_i$  & points with $x$-coordinate at least $x_i$ (\emph{active points}) \\
$\rs_i$ & ranges with at least one corner to the right of $x=x_i$ (\emph{active ranges})\\
$\rs_{i0}$ & subset of $\rs_i$ that are $\tz$ (\emph{active \tz ranges} at $x=x_i$)\\
$P_f$  & \emph{forbidden point set} given by $P_f = P_i \cap (P(Q_0) \cup P(Q_1))$ \\
$\rs_d$  & \emph{deleted range set} given by $\rs_d = \rs_{i0} \cap (\rs(q_0) \cup \rs(q_1))$ \\
\hline
\end{tabular}
\caption{A table of commonly used notations and their explanations.}
\label{table:notation}
\end{center}
\end{table}

\subsubsection{Combining them together}
In the following, we will combine the dynamic programs for \tz and \to ranges to obtain a dynamic 
program for max-exposure in a unit square $C$. We will need a couple of changes.
First, the events at $x=x_i$ are now defined by either a point $p_i \in P$ or beginning of a \to range $R_i$.
Next, the \emph{deleted range set} $\rs_d$ at $x=x_i$ will only consist of \tz ranges and is defined as
$\rs_d = \rs_{i0} \cap (\rs(q_0) \cup \rs(q_1))$ where $\rs_{i0} \subseteq \rs_i$ is 
the set of \tz ranges that intersect the vertical line $x=x_i$.
The \emph{forbidden point set} $P_f = P_i \cap (P(Q_0) \cup P(Q_1))$ stays the same. Here
$q_0, q_1, Q_0, Q_1$ are same as defined before. 
(For the sake of convenience, Table~\ref{table:notation} lists these notations with explanation.)

The subproblems represent the maximum number of points in $P_i \setminus P_f$ that 
can be exposed by deleting $k'$ ranges from $\rs_i \setminus \rs_d$. 
If $k_i = |(\rs(p_i) \cap \rs_{i0}) \setminus \rs_d|$, then we obtain the following combined recurrence.

\begin{align*}
	&S(i,~ k',~q_0,~q_1, Q_0, Q_1)  ~=~   \\
	&\max
	 \begin{cases}
			S(i+1,~ k',~q_0,~q_1,~Q_0,~Q_1) 	   				&~\hspace{-8em}\text{if $p_i \in P_f$, cannot expose $p_i$} \\
			S(i+1,~ k',~q_0,~q_1,~Q_0,~Q_1) 						&~\hspace{-6em}\text{choose to not expose $p_i$}\\
			S(i+1,~ k' - k_i,~\textit{closer}(q_0, p_i), ~\textit{closer}(q_1, p_i),~ Q_0,~ Q_1) + 1  &\hspace{0.3em}\text{expose $p_i$}
	 \end{cases}
	\\
	&\text{\hspace{2em}(\emph{event $x=x_i$ was a point $p_i \in P_i$})}\\ 
	&\max
	\begin{cases}
		S(i+1,~ k'-1,~q_0,~q_1, ~Q_0, ~Q_1) 									    &\hspace{-2.8em}\text{delete \to range $R_i$}\\
		S(i+1,~ k',~q_0, ~q_1, ~\textit{farther}(Q_0, R_i),~\textit{farther}(Q_1, R_i)) 		&\hspace{1em}\text{$R_i$ not deleted}
	\end{cases}
	\\
	&\text{\hspace{2em}(\emph{event $x=x_i$ was beginning of a \to range $R_i \in \rs_i$})} 
\end{align*}

The optimal solution is given by $S(0,~k,~\pbf,~\ptf,~\rbf,~\rtf)$.
The correctness of the above formulation follows from the fact that when we choose to expose $p_i$,
we are guaranteed that all \to ranges in $\rs(p_i)$ have already been deleted, and
the expression $k_i$ only charges for \tz ranges containing $p_i$.
As for the running time, for each event $x=x_i$, we compute $O(kn^2m^2)$ entries and computing 
each entry takes constant time. Since there are $O(n+m)$ events, we obtain the following.
\begin{lemma}
	\label{lemma:restrictedMaxExposure}
	Given a set $P$ of $m$ points in a unit square $C$ and a set of $n$ unit square ranges $\rs$,
	we can compute their max-exposure in $O(k(n+m)n^2m^2)$ time.
\end{lemma}

\subsection{A Constant Factor Approximation}
\label{sec:4-approx}
\begin{figure}[h!]
	\centering
	\includegraphics{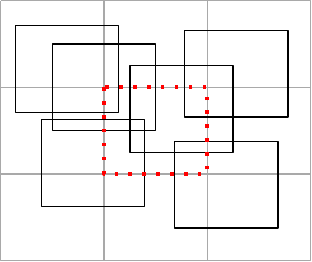}
	\caption{Embedding a max-exposure instance with unit square ranges on a unit-sized grid. 
					 Optimal solution in each grid cell can be computed exactly using Lemma~\ref{lemma:restrictedMaxExposure}.}
	\label{fig:unit-rect.pdf}
\end{figure}

We now use the preceding algorithm to solve the max-exposure problem for general set
of points and unit square ranges within a factor $4$ of optimum.
In particular, we compute a set of $4k$ ranges in $\rs$ such that the number of points
exposed in $P$ by deleting them is at least the optimal number of points. 
Suppose we embed the ranges $\rs$ on a uniform unit-sized grid $G$, 
and define $\cc$ as the collection of all cells in $G$ that contain at least
one point of $P$. Then we can solve exactly for each cell in $\cc$ and combine them
using dynamic programming as described in Algorithm~\ref{algorithm:dp-approx} (DP-Approx).
See also Figure~\ref{fig:unit-rect.pdf}.

\begin{algorithm}
\begin{enumerate}
	\item Apply Lemma~\ref{lemma:restrictedMaxExposure} to solve max-exposure locally 
				in every cell $C_i \in \cc$ for all $0 \leq k_i \leq k$. Call this a \emph{local} 
				solution denoted by $\textit{local}(P(C_i), \rs(C_i), k_i)$, where $P(C_i) \subseteq P$ 
				is the set of points contained in cell 
				$C_i$  and $\rs(C_i)$ is the set of ranges intersecting $C_i$.
	\item Process cells in $\cc$ in any order $C_1, C_2, \dots, C_g$, and define $\textit{global}(i,~ k')$ 
				as the maximum number of points exposed in the cells $C_i$ through $C_g$ using $k'$ ranges. 
				Combine local solutions to obtain $\textit{global}(i,~ k')$ as follows.
				\begin{align*}
				\textit{global}(i,~ k') ~=~ \max_{0 \leq k_i \leq k'}~ \textit{global}(i+1,~ k'- k_i) ~+~ \textit{local}(P(C_i),~ \rs(C_i),~ k_i)
				\end{align*}
	\item Return $\textit{global}(1, 4k)$ as the number of exposed points.
\end{enumerate}
	\caption{DP-Approx}
	\label{algorithm:dp-approx}
\end{algorithm}

\begin{lemma}
	\label{lemma:4-approx}
	If $P^* \subseteq P$ is the optimal set of exposed points, then $\textit{global}(1, 4k) \geq |P^*|$, that
	is , the algorithm \emph{DP-Approx} achieves a $4$-approximation and runs in $O(k (n+m) n^2m^2)$ time.
\end{lemma}
\begin{proof}
Consider the optimal set of ranges $\rs^* \subseteq \rs$. Observe that each 
range $R \in \rs^*$ intersects at most four grid cells.
Let $R_i = R \cap C_i$ be the rectangular region defined by intersection of $R$ and $C_i$.
Clearly, there are at most four regions $R_i$ for each $R \in \rs^*$ and therefore $4k$  in total.
At this point, the regions in cell $C_i$ are disjoint from regions in some other cell $C_j \in \cc$.
Therefore, optimal solution exposes $|P^*|$ points over a set of cells $\cc^*$ such that the set $\rs^*$
has at most $4k$ disjoint components in the cells $\cc^*$. 
Since we can solve the problem exactly for each cell and can combine them using the above
dynamic program, we have that $\textit{global}(1, 4k) \geq |P^*|$ and we achieve a $4$-approximation.

For the running time, we observe that solving max-exposure locally in a cell $C_i$ takes $O(k(n_i+m_i) n_i^2m_i^2)$
time, where $n_i$ is the number of ranges that intersect $C_i$ and $m_i$ is the number of points in $P$ that
lie in $C_i$. Summed over all cells, we get the following bound.
\begin{align*}
	\sum_i k (n_i+m_i) n_i^2 m_i^2  ~&\leq~  k ~\sum_i (n_i + m_i) ~\sum_i  n_i^2 ~\sum_i  m_i^2 \\ 
~&\leq k(n+m)~ (\sum_i n_i)^2 ~(\sum_i m_i)^2 ~=~ O(k(n+m)n^2 m^2)
	\end{align*}
	Once the local solutions are computed, the dynamic program that merges them into a global solution
	has $O(k|\cc|)$ subproblems and computing each subproblem takes $O(k)$ time. Recall that every cell
	in $\cc$ contains at least one point, so $|\cc| \leq n$ and the merge step takes an additional 
	$O(k^2 n)$ time.
\end{proof}

\subsection{Towards a PTAS}
\input{ptas}

\section{\boldmath Extensions and Applications}
\label{sec:extensions}
In this section, we discuss some extensions and applications of our the results 
from previous section. We say that the range family $\rs$ consists of \emph{fat rectangles}
if every range $R\in \rs$ is a rectangle of bounded \emph{aspect ratio}. Moreover, we say
that $\rs$ consists of \emph{similar and fat rectangles}, if ranges in $\rs$ are
rectangles and the ratio of the largest to the smallest side in $\rs$ is constant.
We show that if $\rs$ consists of \emph{similar and fat} rectangles, one can achieve
a constant approximation. Moreover, if $\rs$ consists of \emph{fat rectangles}
one can achieve a bicriteria $O(\sqrt{k})$-approximation. 

\subsection{Approximation for Similar and Fat Rectangles}
Let $a, b$ be the length of smallest and largest sides of rectangles in $\rs$ such that
$b/a = c$ is constant. Then we can modify the input instance as follows. Replace each
range $R \in \rs$ by covering it with at most $c^2$ squares of sidelength $a$ such that the
areas occupied by $R$ and its replacements are the same. Now, we have a modified 
set of ranges $\rs'$ consisting of squares that have the same sidelength. 
Consider the optimal solution with $k$ ranges $\rs^*$ that exposes $m^*$ points. 
It is easy to see that the set $\rs^*$ corresponds to at most $c^2k$ ranges
in the modified instance, and therefore deleting $c^2k$ ranges from $\rs'$ exposes
at least $m^*$ points. Therefore, we can run the polynomial-time $4$-approximation algorithm 
(Lemma~\ref{lemma:4-approx}) to obtain a set of at most $4c^2k$ ranges that 
expose at least $m^*$ points.
\begin{theorem}
	Given a set of points $P$, a set of rectangle ranges $\rs$ such that the ratio of the largest
	to the smallest side in $\rs$ is bounded by a constant, then there exists a polynomial
	time $O(1)$-approximation algorithm for max-exposure.
\end{theorem}

\subsection{Approximation for Fat Rectangles}
We now consider the case when rectangles in $\rs$ have bounded aspect ratio. That is for
all rectangles $R \in \rs$, the ratio of its two sides is bounded by a constant $c$. We 
transform the input ranges $\rs$ to obtain a modified set of ranges $\rs'$ as follows. 
For each rectangle $R \in \rs$, let $x$ be the length of the smaller side of $R$. Then
we replace $R$ by at most $\lceil c \rceil$ \emph{squares} each of sidelength $x$. If $m^*$ is
the optimal number of points exposed by deleting $k$ ranges from $\rs$, then there exists
a set of $O(k)$ ranges in $\rs'$ deleting which will expose at least $m^*$ points.
Observe that the set $\rs'$ consists of \emph{square} ranges, of possibly different sizes.
Therefore, if we can obtain an $f$-approximation for square ranges, we can easily
obtain $O(f)$-approximation with fat rectangles.

\subsubsection{A Bicriteria $O(\sqrt{k})$-approximation for Squares}
We will describe an approximation algorithm for the case when the set of ranges $\rs$ 
consists of axis-aligned squares. We achieve an approximation algorithm in three steps. 
First, we partition the point set by assigning the points to one of the input squares. 
Next, we solve the problem exactly for a fixed input square.
Finally, we combine these solutions to achieve a good approximation to the optimal solution.

We define $\ca : P \rightarrow \rs$ to be a function that assigns a point in $P$ to 
exactly one range in $\rs$. If $\rs(p_i)$ is the set of squares
that contain $p_i$, then $\ca(p_i)$ is the smallest square in $\rs(p_i)$.
This assignment scheme ensures the following property.
\begin{lemma}
	\label{lemma:biggerSquares}
	Let $R \in \rs$ be a square and let $\cp_R = \ca^{-1}(R)$ be the set of points \emph{assigned} to it.
	Moreover, let $\rs' \subseteq \rs$ be the set of squares that intersect $R$ and contain at
	least one point in $\cp_R$. Then, every square $R' \in \rs'$ must have sidelength bigger than
	that of $R$, and therefore contains at least one corner of $R$.
\end{lemma}

Now suppose we fix a square $R$, and consider a restricted max-exposure instance with
the set of its assigned points $\cp_R$. Since, ranges that contain a point in $\cp_R$
are all bigger then $R$, this case is essentially the same as points inside a unit square, 
and therefore Lemma~\ref{lemma:restrictedMaxExposure} can be easily extended to solve it exactly.
This gives us the following algorithm. Here $1 \leq \alpha \leq k$ is a parameter.
\begin{algorithm}
\begin{enumerate}
	\item For every square $R \in \rs$, apply Lemma~\ref{lemma:restrictedMaxExposure} over the
				point set $\cp_R$ to expose the maximum set of points $P(R, k) \subseteq \cp_R$ by deleting
				$k$ ranges.

	\item Order squares in $\rs$ by decreasing $|P(R, k)|$ values, and pick the set $\cs \subseteq \rs$
				of first $\alpha$ squares. 
	\item Return $\bigcup_{R \in \cs} P(R, k)$ as the set of exposed points.
\end{enumerate}
\caption{Greedy-Squares}
\end{algorithm}

\begin{lemma}
	\label{lemma:k-approx-squares}
	Let $m^*$ be the optimal number of points exposed using $k$ squares, then algorithm \emph{Greedy-Squares}
	computes a set of at most $\alpha k$ squares that expose at least $\alpha m^* / k$ points.
\end{lemma}

\begin{proof}
It is easy to see that the number of squares is at most $\alpha k$. To show the bound on
number of points exposed, consider the optimal set $\rs^*$ of $k$ ranges 
and let the optimal set of points exposed by $\rs^*$ to be $P^*$.
We will now use the same assignment procedure $\ca^* : P^* \rightarrow \rs^*$
to assign points in $P^*$ to a square in $\rs^*$. That is, $\ca^*(p_i)$ is the smallest
square in $\rs^*$ that contains $p_i$. We claim that $\ca^*(p_i) = \ca(p_i)$ for all $p_i \in P^*$
since every square that contains $p_i$ lies in $\rs^*$. Moreover, let $\cp_R^*$ denote the
set of points of $P^*$ assigned to $R$.

Let $m'$ be the number of points exposed
by the algorithm and assume that the squares in $\rs$ are ordered such that
$|P(R_i, k)| \geq |P(R_j, k)|$ for all $i< j$. Then, we have the following.

\begin{align*}
	m^* ~&=~ \left| \bigcup_{R \in \rs^*}  \cp_R^* \right| ~=~ \sum_{R \in \rs^*} |\cp_R^*| \\
			 ~&\leq~ \sum_{1 \leq i \leq k} |P(R_i, k)| 
			 ~\leq~ \frac{k}{\alpha} \sum_{1 \leq i \leq \alpha} |P(R_i, k)| 
			 ~=~ \frac{k}{\alpha} m'
\end{align*} 
\vspace{-1em}
\end{proof}

For $\alpha = \sqrt{k}$, the above algorithm achieves a bicriteria $O(\sqrt{k})$-approximation.
Since an $f$-approximation for square ranges gives an $O(f)$-approximation for fat rectangles, we obtain the
following.
\begin{theorem}
	Given a set of points $P$ and a set of ranges $\rs$ consisting of rectangles of bounded aspect ratio, then 
	one can obtain a bicriteria $O(\sqrt{k})$-approximation for max-exposure in polynomial time.
\end{theorem}

\section{Conclusion}
	In this paper, we introduced the max-exposure problem, proved its hardness, and explored
	approximation schemes for it. We showed that the problem is hard to approximate even when 
	the range space $\rs$ consists of two types of rectangles. 
	When the ranges are defined by translates of a single rectangle,  we presented a polynomial-time
	approximation scheme (PTAS).
	Some natural questions to explore in the future include better approximation algorithms, and
	simpler range spaces such as those defined by axis-aligned squares. 
	For instance, can one achieve a constant factor approximation for axis-aligned squares?

\bibliography{refs}
\end{document}

%% file: ptas.tex
In this section, we will show how to extend the exact algorithm for the restricted max-exposure instance
where all points lie inside a unit square (Lemma~\ref{lemma:restrictedMaxExposure}) to obtain an exact solution 
for the max-exposure instance where all points are contained in a $h \times h$ square $\cc$.
Without loss of generality, we can assume that the lower left corner of $C$ is at the origin $(0, 0)$ and
$\cc$ is subdivided into $h^2$ unit-sized grid cells.

\begin{figure}[h!]
\centering
\includegraphics{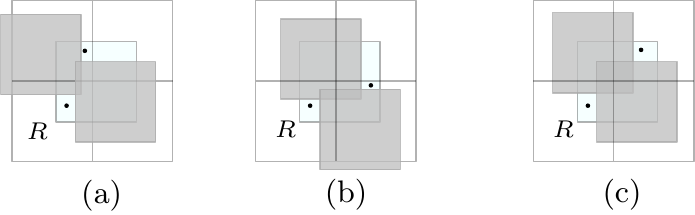}
\caption{Examples where a deleted range $R$ can potentially be counted in two cells that are:~ 
(a)~ vertical neighbors~ (b)~ horizontal neighbors (c)~ diagonal neighbors}
\label{fig:double-counted}
\end{figure}

Observe that a major hurdle in generalizing the dynamic program from Section~\ref{sec:dp-unit-square} for
max-exposure in a unit square cell to the grid $\cc$ is that a range $R$ can be \emph{double counted} in
multiple cells. Specifically, range $R$ may contain exposed points in at most four cells of $\cc$ and  
can be counted in each one of them. (See also Figure~\ref{fig:double-counted}.)
Indeed a natural generalization of the earlier dynamic program to $h$ anchor lines avoids double
counting of ranges in the same column of $\cc$ (vertical neighbors). However, some additional work is required 
to avoid double counting in adjacent columns (horizontal and diagonal neighbors).

To handle this, we first apply the following transformation which we call \emph{flattening} of the grid $\cc$.

\begin{figure}[h!]
\centering
\includegraphics{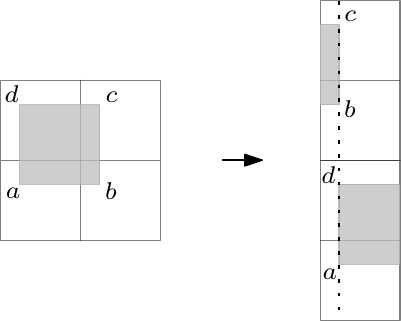}
\caption{Flattening a $2\times2$ grid containing one unit square range that is split into \tz and \to components.}
\label{fig:grid-flatten}
\end{figure}

\paragraph{Flattening the grid $\cc$} Intuitively, the flattening process transforms a $h \times h$ grid into
a $h^2 \times 1$ vertical slab by \emph{shifting} the $i$-th column and aligning it on top of the $(i-1)$-th column. 
More precisely, we label the cells \emph{column by column} from \emph{left to right} and bottom to top in each
column. That is, cells of the column $1$ are labeled as $1, 2, \dots, h$ and the cells of column $2$ are 
labeled $h+1, \dots, 2h$ and so on. Then, flattening refers to simply stacking all the cells in their numbered order.
In other words, we shift the coordinates of all points and parts of ranges in column $i$ of the grid $\cc$ 
by $(-(i-1), (i-1)h)$, for all $2 \leq i \leq h$. (See also Figure~\ref{fig:grid-flatten}.)

After this transformation, all $x$-coordinates are within the range $[0, 1]$ and
$y$-coordinates are within the range $[0, h^2]$. Moreover, every range $R$ is
split into two possibly disconnected half-ranges which preserve the following important
property that follows readily from the fact that the ranges are unit squares.
\begin{lemma}
	\label{lemma:diagonal-neighbor-property}
	Let $R$ be a range and $\cc_{i}, \cc_{i+1}$ be the two consecutive columns
	of the grid $\cc$ intersected by range $R$.
	Then, $R$ is \to with respect to cells in $\cc_i$ and \tz with respect to cells in $\cc_{i+1}$,
	and after the \emph{flattening} transformation, the $x$-coordinate at which $R$ \emph{begins}
	as a \to range in $\cc_i$ is the same as the $x$-coordinate at which $R$ \emph{finishes} as
	\tz range in $\cc_{i+1}$.
\end{lemma}
\begin{proof}
	The range $R$ intersects the vertical line $x=i$ which is coincident with the right (resp. left) boundary
	of cells in $\cc_i$ (resp. $\cc_{i+1}$). Therefore, $R$ is
	\to in cells of $\cc_i$ and $\tz$ in cells of $\cc_{i+1}$.
	
	Let the $x$-coordinate of left boundary of $R$ (that lies in $i$-th column) be $(i-1) + \delta$.
	Therefore, the $x$-coordinate of right boundary of $R$ would be $(i-1) + \delta + 1 = i + \delta$,
	and it will lie in $(i+1)$-th column. After the transformation both these coordinate values would be $\delta$.
\end{proof}

From the above lemma, it follows that every range $R \in \rs$ has a $\tz$ component and a $\to$ component which
may lie in non-consecutive cells. In the rest of the discussion, we will refer to these components by their
type as prefix. For example, $\tz$ range $R$ refers to the $\tz$ component of $R$.

Once we have flattened the grid $\cc$, our algorithm is an almost straightforward extension of the dynamic program 
from Section~\ref{sec:dp-unit-square} to $h^2+1$ anchor lines $\ell_0, \ell_1, \ell_{h^2}$.
Same as before, we process the two types of events : \emph{$x=x_i$ is a point $p_i$} and 
\emph{$x=x_i$ is beginning of \to range $R_i$}. However at every $x=x_i$, we will now 
need to remember the set $\pvec = \{q_0^+, q_0^-, \dots, q_{h^2}^+, q_{h^2}^-\}$ of $O(h^2)$ points 
consisting of \emph{closest exposed points} $q_j^+, q_j^-$ respectively above and below every anchor line $\ell_j$.
Similarly, we will need to remember the set $\rvec = \{Q_0^+, Q_0^-, \dots, Q_{h^2}^+, Q_{h^2}^-\}$ of $O(h^2)$ ranges 
consisting of \emph{farthest undeleted \to ranges} $Q_j^+, Q_j^-$ respectively above and below every anchor line $\ell_j$.

Then at $x=x_i$, we extend the definitions from Table~\ref{table:notation} to obtain the \emph{forbidden point set}
$P_f = P_i ~\cap~ \bigcup_{Q \in \rvec}~ P(Q)$ and the \emph{deleted range set}
$\rs_d = \rs_{i0} ~\cap~ \bigcup_{q \in \pvec}~ \rs(q)$.
Recall that $P_i$ is the set of points with $x$-coordinate at least $x_i$ and $\rs_{i0}$ is the set of $\tz$ ranges
that are active at $x=x_i$. Also recall that
$P(Q)$ denotes the set of points contained in range $Q$ and $\rs(q)$ denotes the set of ranges
containing point $q$. This gives us the following dynamic program which we will
refer to as \textsc{DP-Flattened}.
Same as before, we have $k_i = |(\rs(p_i) \cap \rs_{i0}) \setminus \rs_d|$.
\begin{align*}
	&S(i,~ k',~\pvec,~ \rvec)  ~=~   \\
	&\max
	 \begin{cases}
			S(i+1,~ k',~ \pvec,~ \rvec) 	   				&\hspace{1em}\text{if $p_i \in P_f$, cannot expose $p_i$} \\
			S(i+1,~ k',~ \pvec,~ \rvec) 						&\hspace{1em}\text{choose to not expose $p_i$}\\
			S(i+1,~ k' - k_i,~\closer(\pvec,~ p_i), ~ \rvec) + 1  &\hspace{1em}\text{expose $p_i$}
	 \end{cases}
	\\
	&\text{\hspace{2em}(\emph{event $x=x_i$ was a point $p_i \in P_i$})}\\ 
	&\max
	\begin{cases}
		S(i+1,~ k',~ \pvec,~ \rvec) 	   				&\hspace{4.2em}\text{if $R_i \in \rs_d$, already deleted} \\
		S(i+1,~ k'-1,\pvec,~ \rvec) 									    &\hspace{4.2em}\text{delete \to range $R_i$}\\
		S(i+1,~ k',~ \pvec,~ \farther(\rvec,~ R_i)) 		&\hspace{4.2em}\text{$R_i$ not deleted}
	\end{cases}
	\\
	&\text{\hspace{2em}(\emph{event $x=x_i$ was beginning of a \to range $R_i \in \rs_i$})} 
\end{align*}

Here, $\closer(\pvec, p_i)$ denotes the operation of updating the appropriate closest 
exposed point in $\pvec$ with point $p_i$. More precisely, let $C_j$ be the cell bounded
by anchor lines $\ell_{j-1}$ and $\ell_{j}$ that contains the exposed point $p_i$.
We update $\pvec$ such that $q_{j-1}^+ = \closer(q_{j-1}^+, p_i)$ and $q_{j}^- = \closer(q_{j}^-, p_i)$.
Similarly, let $\ell_j$ be the anchor line intersecting $R_i$, then $\farther(\rvec, R_i)$
denotes the operation of updating $\rvec$ with 
the farthest undeleted range on both sides of $\ell_j$ as 
$Q_j^+ = \farther(Q_j^+, R_i)$ and $Q_j^- = \farther(Q_j^-, R_i)$.
The optimal solution is given by $S(0,~ k, ~\pvecf, ~\rvecf)$, where 
$~\pvecf, ~\rvecf$ consist of the initial values for each anchor line.

At any event $x=x_i$, the above dynamic program accounts for the cost of deleting a range $R$ in one of
two ways: either as a \tz range included in the term $k_i$ or as \to range by paying unit cost.
In the next lemma, we show that every deleted range is counted exactly once and use it to establish
the correctness.
\begin{lemma}
	\label{lemma:slab-polytime}
	The dynamic program \textsc{DP-flattened} computes an optimal solution for max-exposure instance $(\rs, P, k)$
	in an $h \times h$ grid and runs in $O(k(nm)^{O(h^2)})$ time.
\end{lemma}
\begin{proof}
	The running time bound follows from the number of exposed points and undeleted ranges we need to remember.

	To prove correctness, consider an optimal set of deleted ranges $\rs^*$ and its exposed points $P^*$.
	Let the \emph{value} of the solution returned by the dynamic program be the number of points it choses
	to expose and the \emph{cost} of the solution is the total cost of ranges it deletes.
	First, we claim that there exists a sequences of choices at events $x=x_i$ where the dynamic program 
	selects points and ranges consistent with the optimal solution,
	that is, chooses to only expose points in $P^*$ and to only delete \to ranges in $\rs^*$.
	This is easy to verify because because $P^* \cap P_f = \emptyset$, so the dynamic program
	can choose to expose point $p_i \in P^*$ when $x=x_i$ is a point-event.
	Indeed the value of the solution is $|P^*|$. Next we will show that every range in $\rs^*$ is counted
	exactly once, and therefore the cost of the solution is also $k$. 
	
	We claim that at every \emph{point-event} $x=x_i$ where we expose the point $p_i \in P^*$, all
	ranges in $\rs(p_i)$ are deleted and counted exactly once. 
	To see this, let $R \in \rs(p_i)$ be a range containing $p_i$ and let $x=x_r$ be the $x$-coordinate at 
	which $R$ finishes as a \tz range and starts as a \to range. We have two disjoint cases.
	\begin{enumerate}
		\item \emph{$p_i$ is contained in \tz component of $R$.}~ Let $\ell_j$ be the line to which $\tz$ range $R$
			is anchored. We have two subcases.
			\begin{enumerate}
				\item \emph{$R$ does not contains any exposed point to the left of $x=x_i$.} In this case, 
				we charge for $R$ and remember that $R$ has already been counted using the closest exposed points $q_j^+, q_j^-$
				above and below $\ell_j$. Therefore, we will have $R \in \rs_d$ at least until $x=x_r$, 
				(when it switches from being \tz to \to). 
				Since $R$ cannot be charged at $x > x_r$, it is charged exactly once in total.
			
				\item \emph{$R$ contains an exposed point to the left of $x=x_i$.} Then we will have $R \in \rs_d$, 
				and as discussed above $R$ was already counted and would not be charged again.
			\end{enumerate}

		\item \emph{$p_i$ is contained in \to component of $R$.}~ Since $p_i$ is exposed, it is not contained in the
			forbidden point set $P_f$. Therefore, $R$ must be deleted when it began as a \to
			event at $x=x_r$ or else we would have $p_i \in P_f$.
			As discussed above, if $R$ was deleted as a \tz range to the left of $x=x_r$, we must have $R \in \rs_d$
			at $x=x_r$, so it would not be charged again. If $R$ was not deleted as a \tz range, 
			then it would be charged at $x=x_r$ as a \to range and is never charged again.
	\end{enumerate}
	Therefore, the solution returned by dynamic program $S(0, k, \pvecf, \rvecf)$ has
	value at least optimal.
\end{proof}

\subsection{A $(1+\eps)$-Approximation Algorithm}
	We will now apply grid shifting technique by Hochbaum and Maas~\cite{hochbaum1985approximation} 
	to obtain an $(1+\eps)$-approximation\footnote{The PTAS presented here simplifies and corrects an error in the PTAS
	that appeared in the conference version~\cite{max_exposure_2019} of the paper.}.
	In particular, if $P^*$ is the optimal set of exposed points, 
	then we show how to compute a set of $(1+\eps)k$ ranges deleting which will expose at least $|P^*|$ points. 
	Using similar ideas but with small adjustments, we also show how to expose at least $(1-\eps)|P^*|$
	points by deleting exactly $k$ ranges. 

	\begin{theorem}
		There exists an algorithm for max-exposure with unit-square ranges running in $k(mn)^{O(1/\eps^2)}$ time
		that exposes at least optimal number of points by deleting $(1+\eps)k$ ranges.
	\end{theorem}
	\begin{proof}
		For a given shift value $a,b \in \{0, \dots, h-1\}$, we compute the optimal solution inside every
		$h\times h$ cell $\cc_{ij}=  [a+ih,~ a+(i+1)h] \times [b+jh,~ b+(j+1)h]$ for all $i, j \in \mathbb{Z}$.
		Using the exact solution in each cell as \emph{local solution}, we use the algorithm DP-Approx 
		(from Section~\ref{sec:4-approx}) to combine them into a global solution for the entire grid
		given by $S_{ab} = \textit{global}(1, k(1+\eps))$, with $\eps = \lceil 8/h \rceil$.
		We repeat this for every shift $a, b$, and return $S_{ab}$ that achieves the maximum value.

		To see why this exposes at least optimal number of points, consider an optimal
		set of deleted ranges $\rs^*$ and sets $\rs^*_{a}, \rs^*_{b} \subseteq \rs^*$ 
		intersected by \emph{boundary} grid lines $x=a+ih$ and $y=b+jh$ respectively, for all $i,j$.
		These grid lines split the intersected ranges into at most $Z_{ab} = 2|\rs^*_a| + 2|\rs^*_b|$ 
		disjoint components.
		\begin{align*}
		\sum_{0 \leq a, b < h}~ Z_{ab} ~&=~ 
			2h \sum_{0 \leq a < h}~ |\rs^*_{a}| + 2h \sum_{0 \leq b < h}~|\rs^*_{b}| ~\leq~ 8hk \\
			&\implies \min_{0 \leq a, b < h}~ Z_{ab} ~\leq~ 8k/h
		\end{align*}
		The first inequality holds because every range can touch at most two grid lines, so
		we have $\sum_{0 \leq a < h} |\rs^*_{a}| \leq 2k$ and $\sum_{0 \leq b < h} |\rs^*_{b}| \leq 2k$.
		Hence there exists a shift value $a,b$ for which the ranges in optimal solution have at most 
		$k(1+\eps)$ disjoint components in  cells $\cc_{ij}$. 
		Therefore, $\textit{global}(1, k(1+\eps))$ returns at least an optimal number of points.
	\end{proof}

	\begin{theorem}
		There exists an algorithm for max-exposure with unit-square ranges running in $k(mn)^{O(1/\eps^2)}$ time
		that exposes at least $(1-\eps)$ fraction of optimal number of points by deleting $k$ ranges.
	\end{theorem}
	\begin{proof}

		For a given shift value $a,b \in \{0, \dots, h-1\}$, we first \emph{preprocess} the input by discarding points so that the set 
		of ranges intersecting the boundary grid lines $x=a+ih$ and $y=b+jh$ do not contain any point.
		Specifically, for every shift value $a, b$, discard the points that are within a unit distance
		from grid boundary lines $x=a+ih$ or $y=b+jh$ for all $i, j$. On the modified input, 
		we run the exact solution in each cell as \emph{local solution}, and then use DP-Approx 
		(from Section~\ref{sec:4-approx}) to combine them into a global solution for the entire grid
		given by $S_{ab} = \textit{global}(1, k)$.
		We repeat this for every shift $a, b$, and return $S_{ab}$ that achieves the maximum value.

		Let $P^*$ be the optimal set of exposed points. It remains to show that the above algorithm exposes
		at least $(1-\eps)|P^*|$ points. To see this, for the shift value $a, b$, 
		consider the set of discarded points $P^*_a, P^*_b \subseteq P^*$ 
		that are within a unit distance from $x=a+ih$ and $y=b+jh$ respectively. 
		These $|P_a^*| + |P_b^*|$ will not be exposed by our algorithm
		\begin{align*}
		\sum_{0 \leq a, b < h}~ |P_a^*| + |P_b^*| ~&=~ 
			h \sum_{0 \leq a < h}~ |P^*_{a}| + h \sum_{0 \leq b < h}~|P^*_{b}| ~\leq~ 4h|P^*| \\
			&\implies \min_{0 \leq a, b < h}~ (|P_a^*| + |P_b^*|) ~\leq~ 4|P^*| / h
		\end{align*}
		The first inequality holds because every point can lie within a unit distance of at most 
		two horizontal (resp. vertical) lines, so we have $\sum_{0 \leq a < h} |P^*_{a}| \leq 2k$ and 
		$\sum_{0 \leq b < h} |P^*_{b}| \leq 2|P^*|$.
		Therefore, there exists some $a, b$ for which the number of remaining points in the input 
		is at most $(1 - 4/h)|P^*|$. Since every $h \times h$ cell is mutually independent, 
		$\textit{global}(1, k)$ returns at least $(1-\eps)|P^*|$ exposed points, where $\eps = \lceil 4/h \rceil$.
		\end{proof}